\documentclass{article}

\usepackage{arxiv}

\usepackage[utf8]{inputenc} 
\usepackage[T1]{fontenc}    
\usepackage{hyperref}       
\usepackage{url}            
\usepackage{booktabs}       
\usepackage{amsfonts}       
\usepackage{microtype}      
\usepackage{graphicx}
\usepackage{natbib}
\usepackage{doi}
\usepackage{amsmath}

\title{Designing AI for Prosecutorial Governance: Case Prioritization and Statutory Oversight in Mexico}

\author{
  \href{https://orcid.org/0000-0001-8901-6022}{\includegraphics[scale=0.06]{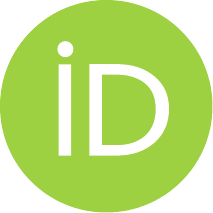}\hspace{1mm}Fernanda Sobrino} \\
  Centro de Ciencia de Datos - EGobiernoyTP \\
  Tecnológico de Monterrey \\
  CDMX, 14380 \\
  \texttt{fersobrinno@tec.mx} \\
  \And
  \href{https://orcid.org/0000-0003-3499-0965}{\includegraphics[scale=0.06]{orcid.pdf}\hspace{1mm}Adolfo De Unánue T.} \\
  Centro de Ciencia de Datos - EGobiernoyTP \\
  Tecnológico de Monterrey \\
  CDMX, 14380 \\
  \texttt{unanue@tec.mx} \\
  \And
  Edgar Hernández \\
  Centro de Ciencia de Datos - EGobiernoyTP \\
  Tecnológico de Monterrey \\
  CDMX, 14380 \\
  \texttt{edgarhr@tec.mx} \\
  \And
  Patricia Villa \\
  Centro de Ciencia de Datos - EGobiernoyTP \\
  Tecnológico de Monterrey \\
  CDMX, 14380 \\
  \texttt{ } \\
  \And
  \href{https://orcid.org/0009-0005-6389-291X}{\includegraphics[scale=0.06]{orcid.pdf}\hspace{1mm}Elena Villalobos} \\
  Centro de Ciencia de Datos - EGobiernoyTP \\
  Tecnológico de Monterrey \\
  CDMX, 14380 \\
  \texttt{villalobos\_elena@tec.mx} \\
  \And
  David Aké \\
  Centro de Ciencia de Datos - EGobiernoyTP \\
  Tecnológico de Monterrey \\
  CDMX, 14380 \\
  \texttt{david.akeuitz@tec.mx} \\
  \And
  \href{https://orcid.org/0009-0000-7625-3782}{\includegraphics[scale=0.06]{orcid.pdf}\hspace{1mm}Stephany Cisneros} \\
  Centro de Ciencia de Datos - EGobiernoyTP \\
  Tecnológico de Monterrey \\
  CDMX, 14380 \\
  \texttt{stephany.cisneros@tec.mx} \\
  \And
  Cristian Paul Camacho Osnay \\
  Fiscal General de Justicia del Estado de Zacatecas \\
  Zacatecas, México \\
  \texttt{jefatura.oficina@fiscaliazacatecas.gob.mx} \\
  \And
  Armando García Neri \\
  Fiscalía General de Justicia del Estado de Zacatecas \\
  Zacatecas, México \\
  \texttt{fisgralzac@outlook.com} \\
  \And
  Israel Hernández\\
  Fiscalía General de Justicia del Estado de Zacatecas \\
  Zacatecas, México \\
  \texttt{israel@fiscaliazacatecas.gob.mx} \\
}

\renewcommand{\headeright}{}
\renewcommand{\undertitle}{}
\renewcommand{\shorttitle}{AI for Prosecutorial Governance}

\hypersetup{
  pdftitle={Designing AI for Prosecutorial Governance: Case Prioritization and Statutory Oversight in Mexico},
  pdfsubject={},
  pdfauthor={Fernanda Sobrino, Adolfo De Unánue T., Edgar Hernández, Patricia Villa, Elena Villalobos, Dabid Aké, Stephany Cisneros, Cristian Camacho, Armando Neri, Israel Hernández},
  pdfkeywords={Artificial Intelligence, Digital Government, Criminal Justice, Algorithmic Governance, Case Prioritization, Public Sector AI, Decision Support Systems, Mexico},
}

\begin{document}
\maketitle

\begin{abstract}
    Prosecutors across Mexico face growing backlogs due to high caseloads and
    limited institutional capacity. This paper presents a machine learning (ML)
    system co-developed with the Zacatecas State Prosecutor’s Office to support
    internal case triage. Focusing on the Módulo de Atención Temprana (MAT)—the
    unit responsible for intake and early-stage case resolution—we train
    classification models on administrative data from the state’s digital case
    management system (PIE) to predict which open cases are likely to finalize
    within six months. The model generates weekly ranked lists of 300 cases to
    assist prosecutors in identifying actionable files. Using historical data
    from 2014 to 2024, we evaluate model performance under real-time
    constraints, finding that Random Forest classifiers achieve a mean
    Precision@300 of 0.74. The system emphasizes interpretability and
    operational feasibility, and we will test it via a randomized controlled
    trial. Our results suggest that data-driven prioritization can serve as a
    low-overhead tool for improving prosecutorial efficiency without disrupting
    existing workflows.
\end{abstract}


\section{Introduction}

Digital-era governance has fundamentally transformed how public institutions
manage information, deliver services, and exercise administrative discretion
\citep{dunleavy2006digital, fountain2001building}. In criminal justice systems,
where decisions carry profound consequences for individuals and communities,
this transformation presents both opportunities and challenges. Algorithmic
systems can help institutions process large caseloads more efficiently, but
they also raise concerns about transparency, accountability, and the
preservation of professional judgment \citep{janssen2016challenges,
  veale2018fairness}. As governments increasingly deploy artificial
intelligence (AI) in high-stakes domains, understanding how to design,
implement, and oversee these systems becomes essential for both researchers and
practitioners \citep{wirtz2019artificial, bullock2020artificial}.

This paper contributes to the growing literature on AI in government by
presenting a case study of algorithmic decision support in a Mexican state
prosecutor's office. We describe the design, implementation, and governance of
a machine learning system developed to address two interrelated institutional
challenges: (1) supporting prosecutors in identifying cases likely to be
resolved in the near term, and (2) providing systematic oversight of cases that
may have exceeded statutory time limits. The system thus serves a dual
function---enhancing operational efficiency while simultaneously enabling legal
accountability. This dual-purpose design reflects a broader lesson for public
sector AI: algorithmic tools need not be limited to efficiency gains but can
also be structured to support institutional oversight and democratic
governance.

Our analysis responds to calls for empirical research on AI implementation in
government contexts outside the Global North.
While much of the existing literature on algorithmic governance draws on cases
from the United States and Western Europe, criminal justice institutions in
Latin America face distinct challenges related to institutional capacity,
digital infrastructure, and legal frameworks. By documenting the design choices
and implementation experience in a Mexican prosecutor's office, we offer
insights that may inform similar efforts in other resource-constrained
settings.

The prosecutorial backlog in Mexico's criminal justice system poses a major
institutional challenge. Across the country, prosecutors' offices are
responsible for overseeing every stage of criminal proceedings—opening
investigations, collecting evidence, promoting alternative solutions, and
pursuing criminal action in court. However, growing caseloads and limited
resources have led to large volumes of unresolved cases. Between 2018 and 2022,
the number of open files nationwide increased by nearly 60\%, with over 2.9
million active investigations reported by the end of 2022
\cite{MexicoEvalua2023, INEGI2023}.

In this project, we partnered with the Fiscalía General de Justicia del Estado
de Zacatecas to design and evaluate a machine learning (ML) system to support
weekly case prioritization. Our aim is not to automate decisions or restructure
workflows, but to assist prosecutors by providing a ranked list of cases that,
based on historical patterns of case progression, are more likely to be resolved
in the near term.

The lists generated by the models serve as a support tool, helping the
prosecutor's office workers keep potentially actionable cases visible and
accessible, without changing institutional procedures or increasing staff
workload. The goal is to enable more informed prioritization while staying
within current operational capacity.

In addition to identifying cases likely to be resolved in the short term, the
system also helps manage the backlog by highlighting cases at the tail of the
distribution. Using the same prediction framework, we generate weekly lists of
approximately 1000 investigations with the lowest estimated probability of
concluding within the next six months. Historical patterns indicate that many of
these cases have already surpassed statutory deadlines for prescription or show
signs of prolonged inactivity.

This paper presents the design and historical validation of this ML-based
prioritization system. We focus on the Módulo de Atención Temprana (MAT), which
handles intake and early-stage filtering. Using data from the Plataforma de
Integración de Expedientes (PIE)—Zacatecas's official case management
database—we train a set of classification models to predict short-term
resolution probabilities and evaluate their performance using precision-focused
metrics aligned with institutional constraints.

We show that ensemble classifiers, especially Random Forests, can reliably
identify cases likely to finalize within a 6-month horizon, offering a
low-overhead decision-support tool for backlog management. The results from
retrospective validations suggest that such models can be feasibly integrated
into prosecutorial workflows to improve case triaging without disrupting
operations.

\section{Institutional Context: The Zacatecas Prosecutor’s Office}

The Fiscalía General de Justicia del Estado de Zacatecas (FGJEZ) is responsible for overseeing the criminal investigation process across the state. In 2021, it began implementing a strategic criminal prosecution policy aimed at reducing procedural delays and improving efficiency. As part of this effort, the office has adopted a structured case management approach that distributes investigations across specialized units and, more recently, has begun integrating predictive tools to support internal prioritization.
At the core of this structure is the “Modelo de Tres Pisos” (Three-Tier Model), which assigns cases to different prosecutorial units based on procedural stage and case complexity:
 \begin{itemize}
 \item \textbf{Tier 1: Módulo de Atención Temprana (MAT)} — Responsible for initial complaint intake, early-stage investigation, and rapid resolution, either through alternative mechanisms or administrative closure.
 \item \textbf{Tier 2: Unidad de Tramitación Masiva de Casos (UTMC)} — Focuses on cases with identified suspects, emphasizing resolution via conditional suspensions or abbreviated procedures.
 \item \textbf{Tier 3: Unidad de Investigación Especializada (UIE)} — Handles complex or sensitive cases requiring full investigation and trial preparation.
 \end{itemize}
Two additional support units work alongside this tiered structure: the \textbf{Órgano Especializado en Mecanismos Alternativos (OEMASC)}, which facilitates out-of-court settlements, and the \textbf{Unidad de Imputado Desconocido (UID)}, which handles cases without identified suspects. The main analysis in this paper focuses on the MAT unit; Appendix~\ref{app:uid} reports results for applying the same methodology to UID.

As of August 2024, the PIE case management system contained over 234,000 investigation files. Roughly 55,000 remained open from previous years. The system registers an average of 1,831 new cases per month, while only 746 are formally closed, and 217 are resolved through alternative mechanisms. These figures, drawn from administrative data covering 2014 to 2024, highlight the backlog challenges faced by entry-point units like MAT, where high inflow and limited throughput generate persistent caseload pressure.
To support internal case review, this paper presents a predictive prioritization system designed for the MAT unit. Each week, the model produces a ranked list of 300 open cases that, based on historical patterns, resemble files that were previously resolved within a six-month window. We share these lists with unit staff to inform daily decision-making, without modifying formal assignments, workloads, or procedural discretion. The objective is not to change how prosecutors work, but to make potentially resolvable cases more visible—supporting more effective triage within existing capacity and institutional constraints.

\section{Data}
\label{sec:data}

The primary dataset used in this analysis comes from the Plataforma de Integración de Expedientes (PIE), the case management system maintained by the Zacatecas State Prosecutor’s Office. Since its implementation in January 2014, PIE has recorded every investigation initiated across all jurisdictions in the state, serving as the official digital ledger of prosecutorial activity. Each case file within PIE contains structured metadata, entered by prosecutors, investigators, and administrative staff through a secure web interface at key procedural milestones.

PIE captures a broad range of information for each investigation. The information includes fundamental details such as the date and location of the reported offense, the individuals involved, applicable legal classifications, and a brief narrative summary of the alleged events. Additionally, the system logs a chronological sequence of procedural events, including arrests, evidence submissions, protective measures, unit transfers (with reasons and timestamps), court referrals, and final case outcomes (e.g., closures or jurisdictional transfers). While the complete paper dossiers remain offline, PIE preserves the structured fields necessary to reconstruct the full procedural history of each case.

To support data integrity, the PIE IT team performs quarterly audits comparing digital entries to physical case files, and unit supervisors conduct weekly spot checks to monitor for delayed or missing updates. Despite these governance practices, two important limitations remain. First, the current PIE system does not digitize or allow uploading of narrative documents and evidentiary attachments—which could provide deeper insights into case complexity. Second, staff sometimes enter updates in batches, leading to timestamp noise of up to one week in some event records. To address this, we consistently rely on the update timestamps recorded in PIE rather than auxiliary time indicators to preserve consistency in event sequencing.

\section{Features}

Using data from the Plataforma de Integración de Expedientes (PIE), we construct a set of features designed to capture how the Prosecutor's Office interacts with each case over time. Our unit of prediction is the individual case record, specifically those that remain open in the Módulo de Atención Temprana (MAT) at the time of analysis. The features span both static characteristics of the case and dynamic indicators of prosecutorial activity, reflecting not only the nature of the crime but also the administrative and legal progression of the case within the office. This strategy aligns with recent work emphasizing event-driven features and longitudinal context in high-stakes domains \cite{Chouldechova2018, Kleinberg2018}.

We organize the features into five main groups:

\begin{enumerate}
    \item \textbf{Case-Level Characteristics}: These features describe the fundamental attributes of each case:
    \begin{itemize}
      \item Opening details: date the case was opened, principal crime category (categorical), and whether the case began with an arrested suspect (typically when the police initiate the case with a detained individual, often caught in flagrante).
      \item We define a procedural event as any logged activity reflecting meaningful progress or change in a case's status. PIE tracks the following core event types: case initialization, progress updates in the investigation, transfers between units or jurisdictions, issuance or execution of search warrants, updates to suspect information, updates to the involved party information, and case closure or resolution.
      \item Event dynamics: Using these event types, we construct cumulative counts of total events per case, as well as temporal aggregates over the past 3 months, 6 months, 1 year, and 2 years. We also calculate average event rates per period.
      \item Temporal gaps: number of days between the date of the crime and the date it was reported, days since the last recorded event, days the case has remained with the same lawyer, and the total duration since the case was opened.
    \end{itemize}

    \item \textbf{Investigation Milestones}: This group captures key investigative and judicial actions that signal progression or resolution pathways within a case. These include binary indicators and cumulative counts of events such as judicialization, imposition of preventive detention, requests for judicial authorization, issuance of arrest warrants, referrals to conciliation, linkage to trial (vinculación a proceso), and the use of alternative justice mechanisms. For each milestone type, we compute both historical totals and time-bounded aggregates over recent periods, enabling temporal sensitivity in predictive patterns.

    \item \textbf{Lawyer Activity and Performance}: To reflect the prosecutorial dimension of each case, we engineer features that measure both individual lawyer activity and their broader engagement patterns. We record how many different prosecutors have handled a case, and for each assigned lawyer, we calculate their active caseload at the time of the case's most recent event. Historical activity metrics include total opened, closed, and transferred cases, along with productivity ratios such as the share of open cases that have been successfully closed. We compute these over the last month, 3 months, 6 months, and full case history. We also include relative performance rankings among prosecutors, enabling the model to adjust for workload heterogeneity across staff.

    \item \textbf{Unit-Level Indicators}: In parallel with lawyer-specific features, we include metrics that capture performance and workload at the unit level. These include total open and closed cases handled by each unit, recent closure rates, and activity-based rankings relative to other units. These indicators provide a contextual backdrop for understanding how institutional bottlenecks or momentum may shape case trajectories.

    \item \textbf{Crime-Type Aggregates}: Finally, we include features summarizing historical trends within each crime type. For each primary offense category, we compute historical and recent counts of open, closed, and total cases, average event volumes, and closure ratios. Rank-ordering these categories by resolution activity further allows the model to recognize patterns in how different types of crime are typically processed over time.

\end{enumerate}

This feature set reflects principles from interpretable machine learning design, prioritizing temporally-grounded and institutionally meaningful variables \cite{Lakkaraju2016, Molnar2020}. By focusing on longitudinal event structures rather than static case attributes, the model captures not just what kind of case is being handled, but how it is evolving—by whom, for how long, and with what apparent momentum.

\section{Methodology}
\label{sec:methodology}

We frame the problem as a binary classification task: given an open case in the target prosecutorial unit (MAT), predict whether the case will be finalized within the next six months. The aim is to support the unit in prioritizing cases that are most likely to leave its jurisdiction—either because no further prosecutorial action is feasible or because the case is transferred to another unit for continued processing. Model training uses historical case data from across the entire prosecutor’s office, allowing the classifier to learn patterns of case progression observed throughout the institution. Predictions and prioritization are restricted exclusively to cases currently handled by the MAT unit.

To simulate real-time deployment, we use a rolling temporal cross-validation strategy \cite{Roberts2017}. At each prediction point, we train the model only on data available up to that moment, ensuring no information leakage into the future. This approach has been recommended for high-stakes public sector applications where real-time decision support must mirror deployment conditions.

Model training uses case histories from 2014 to 2024, with predictions and evaluation focused on the 2023–2024 period. We evaluate models using Precision@300 and Recall@300, which are operationally relevant metrics aligned with institutional constraints, following similar practices in ML-for-policy research \cite{Lakkaraju2016, jung2017simple}.

We compare several classification models, including Logistic Regression, Decision Trees, Random Forests, and Extra Trees. In addition to these learning algorithms, we use two points of comparison to contextualize model performance: (1) a random selection of 300 cases, and (2) a base rate ranking derived from historical closure probabilities—i.e., the empirical likelihood that similar cases have been finalized in the past. While the first serves as a naive baseline, the second provides a data-informed benchmark against which to assess whether the machine learning models learn patterns beyond historical averages (see Appendix~\ref{app:baseline}).

Overall, this methodology aligns predictive modeling with institutional practice—supporting real-time decision-making, avoiding temporal leakage, and evaluating performance through actionable, interpretable metrics.

A key design principle underlying this system is that algorithmic tools in
government should serve multiple governance functions when possible. Rather
than building separate systems for productivity enhancement and legal
oversight, we intentionally designed a unified framework that addresses both
needs from a single prediction model. The same probability estimates that
identify cases likely to be resolved in the near term also reveal, at the
opposite tail of the distribution, cases that appear stalled and may warrant
administrative review. This dual-purpose architecture reflects an emerging
principle in public sector AI: that well-designed algorithmic systems can
simultaneously support operational efficiency and institutional accountability
\citep{peeters2018digital, veale2018fairness}.

In addition to the primary task of predicting short-term case resolution, we use the same prediction outputs to identify cases at high risk of statutory prescription. The need to facilitate administrative and legal oversight of long-inactive cases that are unlikely to make progress under current circumstances drives this extension. However, it may still require a formal review to determine whether these cases should be closed due to statutory prescription. By using the same model, we can maintain a consistent ranking of cases based on their anticipated future outcomes. This approach allows us to prioritize near-term actions while also assessing long-term dormancy, all derived from a single, coherent representation of case dynamics. Exact legal computation of prescription is not possible with the available administrative data, as the database lacks the legally necessary qualifiers to determine the precise offense subtype, the legally relevant starting point of the limitation period, and formal procedural acts that interrupt the limitation period. We therefore construct an operational approximation grounded explicitly in the prescriptive rules of the Código Penal para el Estado de Zacatecas (CPZac, arts.\ 93–112).

Across the dataset, investigations fall into 183 unique crime categories. However, under CPZac, each category typically contains multiple legally defined subtypes with distinct penalty ranges and corresponding statutory prescription periods. For example, theft may vary by the monetary amount, the use of violence, and aggravating factors, each of which implies a different prison sentence and therefore a different limitation period. To address this legal heterogeneity, we compute statutory prescription periods directly from the penalty structure of CPZac using the legally defined rules:

\begin{itemize}
\item If the offense is punishable only by a fine, the prescription period is set to 1 year (art.\ 97 CPZac).
\item If the offense is punishable only by disqualification, suspension, or deprivation of rights, the prescription period is set to 2 years (art.\ 99 CPZac).
\item If the offense carries a prison sentence (either exclusively or), the prescription period equals the arithmetic mean of the minimum and maximum prison term (\emph{término medio aritmético}, TMA), with a statutory lower bound of 3 years (art.\ 98 CPZac).
\item Certain sexual offenses involving minors or legally incapable victims are legally imprescriptible (art.\ 96, second paragraph CPZac) and are therefore excluded from prescription screening.
\end{itemize}

Because each crime category $c$ may map to multiple legal subtypes with different statutory penalties, we construct three alternative statutory prescription thresholds for each category. Let $\mathcal{S}(c)$ denote the set of all legally defined subtypes associated with crime $c$, and let $T(s)$ denote the statutory prescription period implied by subtype $s \in \mathcal{S}(c)$ under the rules above. We then define:
\[
T_{\min}(c) = \min_{s \in \mathcal{S}(c)} T(s), \quad
T_{\text{mean}}(c) = \frac{1}{|\mathcal{S}(c)|} \sum_{s \in \mathcal{S}(c)} T(s), \quad
T_{\max}(c) = \max_{s \in \mathcal{S}(c)} T(s).
\]

The maximum threshold corresponds to assuming the legally most severe subtype within the category (longest prison sentence), the minimum threshold corresponds to the least penalized subtype, and the mean threshold represents an intermediate statutory approximation across all lawful variants.

For each case $i$ associated with crime $c_i$, with opening date $t_i$ and evaluation date $t$, we flag a case as potentially prescribed under threshold $k \in \{\min, \text{mean}, \max\}$ if:
\[
t - t_i \geq T_k(c_i).
\]

At each weekly prediction point, we flag cases as potentially prescribed when they simultaneously fall within the bottom tail of the predicted resolution probability distribution (lowest-ranked cases) and exceed these statutory time thresholds. This procedure does not aim to establish legal prescription with certainty, as it does not reconstruct interruptive acts or tolling events (arts.\ 102–104 CPZac), but rather to identify files with a high likelihood of having surpassed legally meaningful time horizons under plausible statutory assumptions. This approach narrows administrative review to cases that are both unlikely to progress and plausibly affected by statutory time limits, rather than determining prescription outcomes directly.

In the main Results section, we report estimates based on the mean statutory threshold $T_{\text{mean}}$ as our primary specification. This choice provides a balanced approximation that avoids the overly conservative assumptions of the maximum threshold and the potentially permissive assumptions of the minimum threshold. We report results based on the full range of thresholds in the Appendix as robustness checks.

\section{Results}

Unless otherwise noted, all results in this section refer to cases handled by the MAT unit. Appendix~\ref{app:uid} summarizes experiments for the Unidad de Imputado Desconocido (UID).

\begin{figure}[h]
\centering
\includegraphics[width=\linewidth]{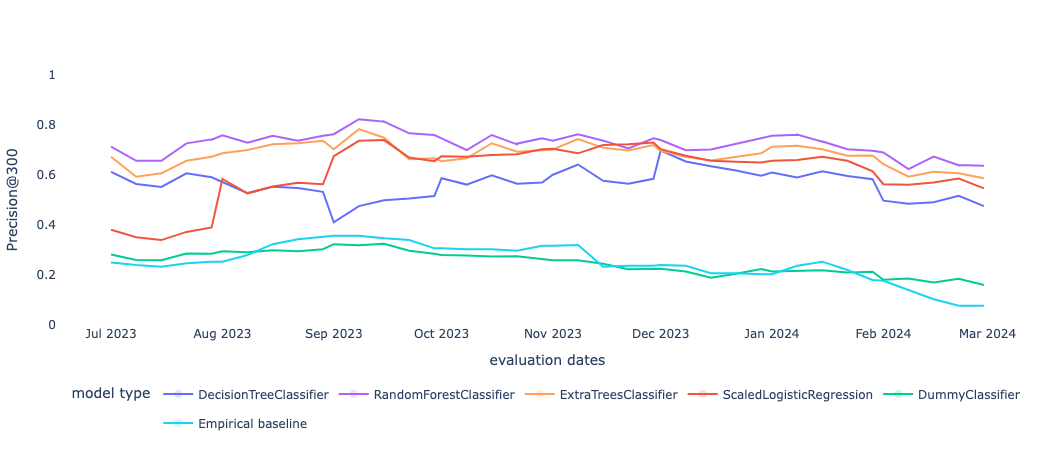}
\caption{Precision@300 for best model per algorithm across evaluation periods.}
\label{fig:best_models}
\end{figure}

Figure~\ref{fig:best_models} presents model performance over time, showing that ensemble methods—particularly Random Forests—consistently outperform simpler baselines. This aligns with findings from other ML applications in legal and public service domains, where ensemble classifiers have proven effective at detecting latent patterns in administrative data \cite{berk2018fairness,kaminski2020predictability}.

As shown, both the Random Forest and Extra Trees classifiers consistently outperform other algorithms over time. While their performance is closely aligned, the Random Forest slightly edges out the Extra Trees on average, making it the top-performing algorithm overall. Simpler models like the Decision Tree and Scaled Logistic Regression also perform reasonably well, but with greater variability across time. The Dummy Classifier, which randomly selects 300 cases, provides a reference point for how much signal the predictive models capture beyond chance. In contrast, the empirical baseline—constructed by ranking cases by historical closure rates, grouped by principal crime, and estimated from the full prosecutor’s office data—offers a data-driven benchmark that reflects past institutional patterns rather than random selection.

\begin{figure}[h]
\centering
\includegraphics[width=\linewidth]{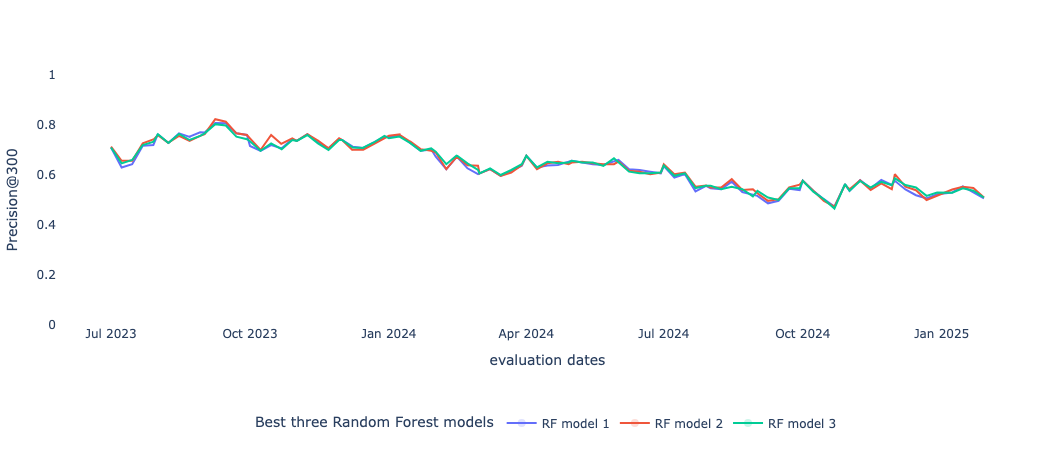}
\caption{Out-of-time Precision@300 of the best-performing Random Forest models evaluated on data through January 2025.}
\label{fig:rf_2025}
\end{figure}

Beyond retrospective evaluation, we evaluated the three best-performing Random Forest models on newly available data extending through January 2025 using the same rolling temporal framework. Figure~\ref{fig:rf_2025} reports the Precision@300 of these models over the extended horizon. Performance remains stable relative to prior evaluation periods, indicating that the models generalize well under real-time conditions and sustain their ability to surface high-likelihood cases over time.

Table~\ref{tab:model_averages} summarizes the average Precision@300 for the best model of each algorithm over the full evaluation window.

\begin{table}[h]
    \caption{Average Precision@300 for the best model per algorithm across evaluation periods}
    \centering
    \begin{tabular}{ll}
        \toprule
        \textbf{Model Type} & \textbf{Average Precision@300} \\
        \midrule
        Random Forest Classifier & 0.735 \\
        Extra Trees Classifier & 0.691 \\
        Scaled Logistic Regression & 0.618\\
        Decision Tree Classifier & 0.573 \\
        Dummy Classifier & 0.258 \\
        Empirical Baseline & 0.253\\
        \bottomrule
    \end{tabular}
    \label{tab:model_averages}
\end{table}

Overall, these results suggest that ensemble-based classifiers—particularly Random Forests—are most effective at identifying high-likelihood case completions under real-time constraints. The consistent gap between these models and both the simpler classifiers and the random baseline highlights the utility of learning from detailed prosecutorial event data.

\subsection{Feature Importance}
\begin{figure}[h]
 \centering
 \includegraphics[width=\linewidth]{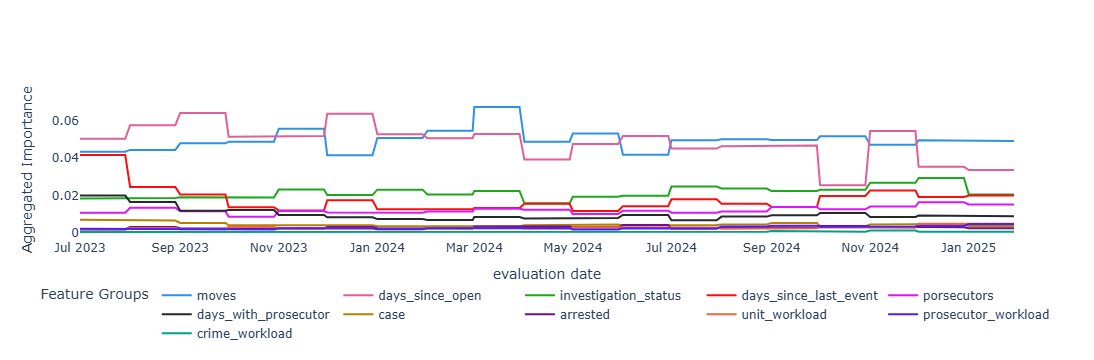}
 \caption{Aggregated feature importance over time by feature group for Random Forest models.}
 \label{fig:feature_groups}
 \end{figure}

To understand what drives the model's predictions, we analyzed feature importances from the top-performing model group: the Random Forest classifiers. Figure~\ref{fig:feature_groups} shows the aggregated importance scores for different feature groups, averaged across evaluation periods. We compute these group-level importances by summing the normalized importance of all individual features within each group, providing a higher-level view of where predictive power resides.

 As shown, the most influential group of features corresponds to procedural movements (moves), which include event-level information such as the number and timing of case status changes, unit transfers, or investigation milestones. This category consistently dominates across all periods, suggesting that case activity patterns offer strong predictive signals of whether a case is nearing resolution. Other important groups include \texttt{days\_since\_open} (how long the case has been active), \texttt{days\_since\_last\_event} (recency of prosecutorial action), and \texttt{estados\_investigacion} (status transitions in the investigation).

\begin{figure}[h]
 \centering
 \includegraphics[width=\linewidth]{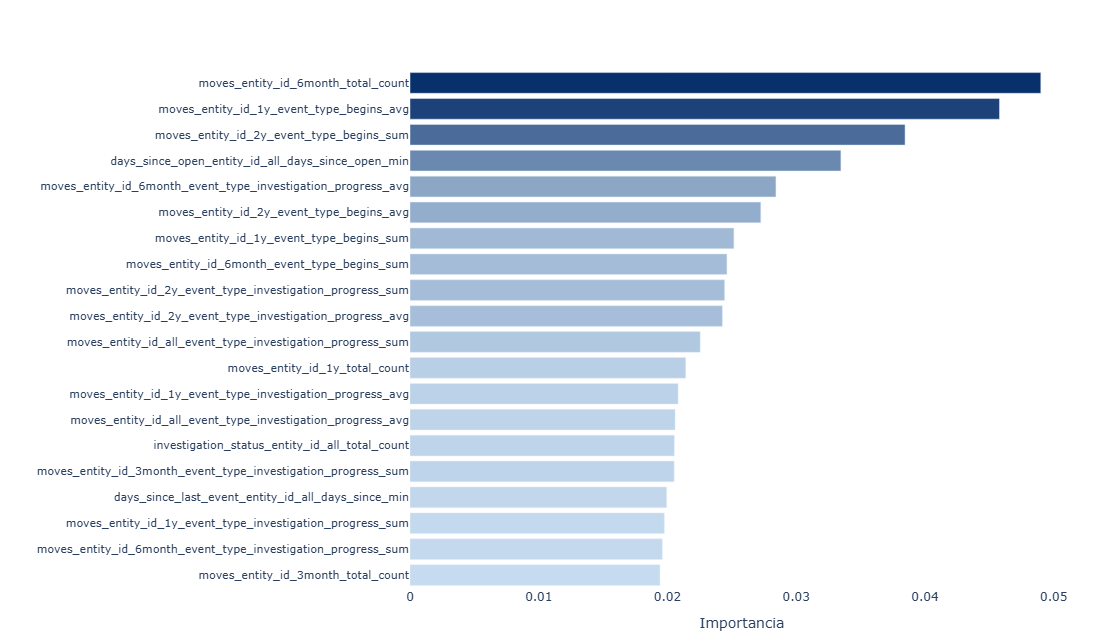}
 \caption{Top 20 most important features in the best-performing Random Forest models.}
 \label{fig:feature_detail}
 \end{figure}

 To drill down into the specific variables contributing most to prediction,
 Figure~\ref{fig:feature_detail} presents the top 20 individual features ranked
 by importance. The strongest single predictor is the number of "initiated"
 events in the past year, followed closely by the same feature measured over
 six-month and two-year windows. In this context, "initiated" refers not to a
 case being reopened after closure, but rather to a formal re-initiation event
 caused by a transfer between units, which triggers a new entry in the PIE
 system. These events reflect administrative handoffs, signaling that a case has
 been reassigned and is now actively progressing under a different team. The
 frequency of these transitions may indicate cases that are moving through the
 system toward resolution.

 Other high-importance features include the minimum number of days since case
 opening, as well as rolling averages of specific event types such as
 \texttt{avance\_investigacion} (investigation progress) and
 \texttt{cambio\_unidad} (unit transfers). These temporal dynamics collectively
 suggest that recent and sustained prosecutorial activity—especially related to
 reassignment and procedural advancement—are key indicators of eventual case
 completion.

 The model primarily depends on dynamic signals that capture how cases
 evolve—particularly the frequency and timing of administrative or investigative
 actions. In contrast, static attributes such as the presence of an arrested
 suspect at intake or generalized crime type statistics contribute less to
 prediction. These results underscore the importance of the longitudinal event
 structure captured in PIE and demonstrate the value of modeling prosecutorial
 workflows over time, rather than relying solely on initial case
 characteristics.

\subsection{Oversight Function: Identifying Potentially Prescribed Cases}

In addition to prioritizing cases with a high likelihood of near-term
resolution, we analyze the opposite tail of the prediction distribution by
examining the 1000 cases with the lowest predicted probability of concluding
within six months. We evaluate these cases for potential statutory prescription
using the three operational definitions described in the Methodology. In this
section, we report results based on the mean statutory prescription threshold
$T_{\text{mean}}$, while we present the minimum and maximum thresholds in
Appendix~\ref{app:prescription_bounds} as robustness checks.

Table~\ref{tab:prescription_mean_share} reports the mean number and
corresponding share of potentially prescribed cases among the bottom 1000 for
the best-performing model of each algorithm under the mean statutory threshold.
Under this specification, the Random Forest model flags, on average, 318.6 cases
as potentially prescribed, corresponding to 31.9\% of the low-priority tail. The
Extra Trees and Scaled Logistic Regression models identify approximately
212--213 cases (about 21.3\%), while the Dummy Classifier flags 230.3 cases
(23.0\%).

\begin{table}[h]
    \caption{Potentially prescribed cases among the 1{,}000 lowest-ranked cases (mean statutory threshold)}
    \centering
    \begin{tabular}{lcc}
        \toprule
        \textbf{Model} & \textbf{Mean Count} & \textbf{Share (\%)} \\
        \midrule
        Random Forest & 318.63 & 31.9 \\
        Extra Trees  & 215 & 21.5 \\
        Scaled Logistic Regression  & 213.43 & 21.3 \\
        Dummy Classifier & 50.95 & 5.1\\
        \bottomrule
    \end{tabular}
    \label{tab:prescription_mean_share}
\end{table}

\begin{figure}[h]
 \centering
 \includegraphics[width=\linewidth]{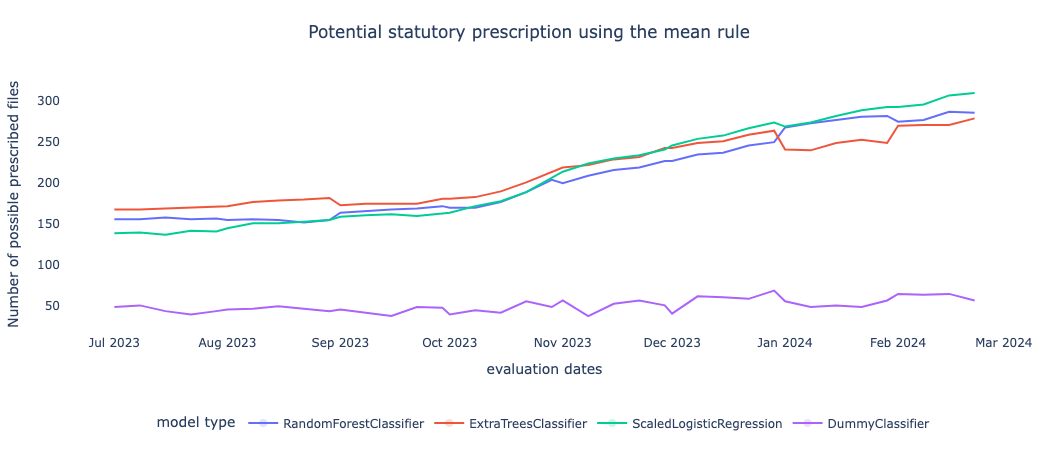}
 \caption{Mean number of potentially prescribed cases among the bottom 1000 cases under the mean statutory threshold, by model family.}
 \label{fig:prescription_trend_all}
\end{figure}

Figure~\ref{fig:prescription_trend_all} shows the temporal evolution of the mean
number of potentially prescribed cases among the bottom-ranked files under the
mean statutory threshold for all model families. Across all model types, the
number of flagged cases exhibits a persistent upward trend over time. This
pattern indicates that prescription risk in the low-priority tail is not
episodic but instead accumulates steadily as cases remain inactive.

Importantly, the trend is qualitatively similar across the Random Forest, Extra
Trees, and Scaled Logistic Regression models, suggesting that the observed
increase reflects structural backlog dynamics rather than model-specific trends.
The Dummy Classifier also displays a rising trajectory, reinforcing the
interpretation that prescription accumulation is driven primarily by
institutional time dynamics rather than predictive ranking alone.

\begin{figure}[h]
 \centering
 \includegraphics[width=\linewidth]{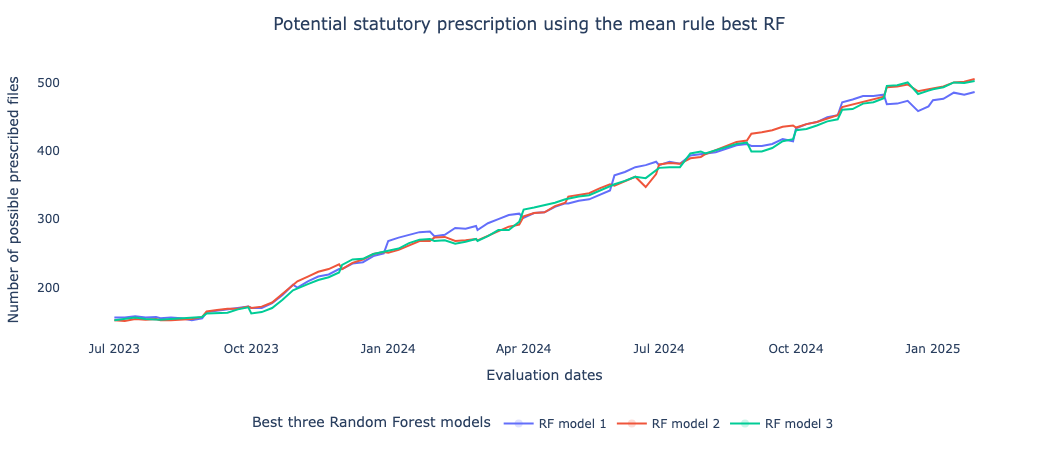}
 \caption{Mean number of potentially prescribed cases among the bottom 1{,}000 using the mean statutory threshold for the three best Random Forest models.}
 \label{fig:prescription_trend_rf}
\end{figure}

To assess robustness within the top-performing Random Forest family,
Figure~\ref{fig:prescription_trend_rf} reports the same prescription dynamics
for the three best Random Forests. Across these configurations, the prescription
signal is highly stable: the mean number of potentially prescribed cases in the
bottom 1000 remains tightly concentrated around 316-319 cases over time. This
stability indicates that the prescription identification is not sensitive to
hyperparameter variation within the Random Forest class.

Taken together, these results indicate that roughly one-third of the
lowest-priority cases identified by the best-performing models are likely to
have surpassed legally meaningful statutory time limits under the mean
prescription rule. This additional list of the lowest-ranked cases complements
the institutional function of the prioritization system. While the top of the
ranking supports short-term prosecutorial productivity, the bottom supports a
systematic, data-driven mechanism for identifying files that warrant
administrative review, formal closure, or corrective action.

\section{Implementation and Future Steps}

While the selected model shows strong performance on historical data, we need to
validate its effectiveness in a real-time operational setting before integrating
it into prosecutorial workflows. To understand the real-world impact of the
model and the prioritization system it supports, we plan to conduct a randomized
controlled trial (RCT) in the Módulo de Atención Temprana (MAT). This evaluation
aims to determine whether prioritizing cases based on the model's output leads
to measurable improvements in prosecutorial throughput under routine operating
conditions \cite{Kleinberg2018}.

The selected model—a Random Forest classifier—achieved an average Precision@300
of 0.74 across historical validation periods. While this indicates strong
discriminatory performance, retrospective accuracy alone cannot establish
real-time utility. We structure the RCT to causally evaluate whether
model-informed prioritization increases the number of cases finalized or reduces
time to resolution relative to the status quo \cite{Chouldechova2018}.

In the proposed experimental design, we will run the model weekly to score all
open cases in the MAT unit. We will identify the 600 cases with the highest
predicted probability of being finalized within the next six months. From this
ranked set, we will randomly assign 300 cases to the treatment group and mark
them for prioritized review, while the remaining 300 will form the control group
and proceed through the standard workflow. Prosecutors will not be informed of
group assignments, and we will track case outcomes using unique administrative
identifiers.

To avoid repeated exposure and dependence across weekly cohorts, we enroll each
case at most once: if a case selected in a prior week remains open and reappears
in the top-ranked set, we exclude it from re-randomization and replace it with
the next-highest-ranked eligible case, ensuring each weekly cohort contains 600
unique cases.

We will conduct the experiment over six consecutive weeks, yielding
approximately 1800 treatment cases and 1800 control cases. We will follow all
cases for six months and compare outcomes along operationally relevant
dimensions such as resolution rates and time to completion
\cite{Chouldechova2018}. This design tests the effect of prioritized
visibility itself, rather than any procedural or staffing change.

In parallel with the productivity-focused RCT, we plan a second evaluation track
focused on the administrative and legal implications of the system's
prescription-screening component. Each week, the model identifies a subset of
low-priority cases that exceed the mean statutory prescription threshold.
Although this signal does not constitute a formal legal determination, it
provides a data-driven list of files warranting administrative review.

Because there is no single outcome metric analogous to resolution for
prescription, evaluation will rely on process-based indicators. Specifically, we
will track the actions taken following the recommendation (e.g., formal
verification of prescription, decision not to prosecute archival, reopening due
to new information, or continued investigation), the time elapsed between the
recommendation and the administrative decision, and the share of reviewed cases
in which the prescription is formally confirmed. This will allow us to assess
whether the system facilitates earlier identification and handling of legally
dormant files, even without a single summary performance statistic.

If both the prioritization and prescription components demonstrate operational
value, we may incorporate the system into the MAT unit's routine workflow with
two complementary outputs: a high-priority list for near-term prosecutorial
action and a low-priority list for administrative and legal review.

Finally, the model architecture and feature design developed in this study are
tightly coupled to Zacatecas's institutional processes, digital infrastructure,
and historical data availability. As a result, the model is not directly
transferable to other jurisdictions without substantial adaptation. Differences
in case management systems, event taxonomies, legal procedures, and digitization
practices would require redesigning both the features and the validation
strategy. Nevertheless, the broader methodological framework—framing triage and
legal dormancy as prediction problems and validating them through policy-aligned
experiments—may be portable to other prosecutorial settings with compatible
administrative data.

\section{Discussion}

This case study offers several insights for the design and governance of AI
systems in public sector contexts. First, the dual-purpose architecture
demonstrates that algorithmic tools can serve both operational efficiency and
institutional oversight objectives simultaneously. By designing a single
prediction framework that supports near-term prioritization at one tail of the
distribution and statutory review at the other, we avoided the need for
separate systems while ensuring that both governance functions receive
attention. This approach aligns with calls for AI systems that embed
accountability mechanisms from the design stage rather than treating oversight
as an afterthought \citep{veale2018fairness, peeters2018digital}.

Second, the implementation experience highlights the importance of
institutional fit. The system was designed around existing workflows rather
than requiring organizational restructuring. Weekly lists are delivered through
channels already familiar to prosecutors, and the ranked format aligns with how
staff already conceptualize case prioritization. This ``low-overhead''
approach, which aims to enhance rather than replace professional judgment,
appears essential for adoption in resource-constrained settings where staff
have limited capacity to learn new systems or adapt to changed procedures.

Third, the case illustrates how administrative data infrastructures shape the
possibilities for algorithmic governance. The Plataforma de Integración de
Expedientes (PIE) provides the digital foundation that makes this kind of
analysis possible, but its structure also constrains what can be predicted. The
system captures procedural events but not the substantive content of case
files; it records timestamps but with some noise; it classifies crimes at a
level of aggregation that introduces uncertainty into prescription
calculations. These limitations are not unique to Zacatecas---they reflect the
broader reality that algorithmic systems in government must work with
administrative data designed for other purposes \citep{janssen2016challenges}.

Finally, this work contributes to the emerging literature on AI in government
contexts outside the Global North. The challenges faced by the Zacatecas
prosecutor's office---high caseloads, limited staff, persistent backlogs---are
common across Latin American criminal justice systems and in many other
developing country contexts. By documenting both the potential and the
limitations of algorithmic decision support in this setting, we hope to inform
future efforts to design AI systems that are appropriate for the institutional
realities of diverse governance contexts.

\section{Ethical Considerations and Limitations}

We designed the prioritization model presented in this paper strictly as a
decision-support tool, not as a substitute for prosecutorial judgment. It does
not alter decision-making authority, interfere with case assignment, or impose
mandatory action thresholds. Instead, it surfaces cases that resemble past
resolutions under existing institutional behavior to assist internal triage
within capacity constraints. Prosecutors retain full discretion over whether and
how to act on model outputs.

However, any model trained on historical administrative data may reflect
patterns of past bias or structural inequality. For example, the prioritization
logic may implicitly favor case types, jurisdictions, or procedural pathways
that have historically moved through the system more quickly. This could
unintentionally reinforce disparities if not carefully monitored
\cite{Barabas2018, Kleinberg2018}.

The prescription-screening component introduces additional ethical and legal
sensitivities. The system does not determine legal prescription, nor does it
reconstruct interruptive procedural acts or tolling events required for a formal
legal declaration. Instead, it produces a risk-based shortlist of cases that
appear to exceed statutory time horizons under operational legal approximations.
There is a nontrivial risk that such signals could be misinterpreted as
definitive legal conclusions if safeguards are not maintained. For this reason,
we frame all prescription-related outputs as recommendations for administrative
review, not as binding legal determinations.

To mitigate broader algorithmic risks, we implemented the system in a
non-intrusive manner: prosecutors retain complete discretion, and no decisions
are made automatically. Moreover, the ongoing randomized controlled trial will
allow us to test whether model-guided prioritization has differential impacts
across subgroups.

We conduct an initial fairness assessment to evaluate whether the model
systematically prioritizes certain types of crime or unintentionally
deprioritizes cases across municipalities or population groups. The analysis
examines subgroup precision, recall, and exposure rates, together with potential
disparities in prioritization and prescription flagging. In preliminary fairness
diagnostics, we examine subgroup performance using historical prediction runs.
At the municipal level, exposure and Precision@300 vary across jurisdictions:
the reference municipality (Zacatecas), which also accounts for the largest
share of cases, exhibits comparatively stable parity, while smaller
municipalities show greater variability. By contrast, crime-type analyses look
more balanced overall, with most categories tracking the reference offense
(simple theft) closely, and clearer deviations appearing mainly in
property-damage cases. These patterns likely indicate that differences in
caseload structure and institutional processing are responsible, but they also
underscore the need to monitor fairness rather than assume it empirically.

We will continue expanding these diagnostics in future iterations, incorporating
additional fairness metrics and stability checks as more deployment data becomes
available.

Finally, because we approximate statutory prescription using generic crime
categories rather than legally exact offense subtypes, all prescription-related
estimates must be interpreted as upper- and lower-bound risk indicators rather
than precise legal conclusions. This limitation reinforces the need for human
legal verification in all downstream uses of the system.

\section*{Data Availability}

The data used in this study contains sensitive criminal justice information and
cannot be publicly shared due to legal and privacy restrictions under Mexican
law. The administrative records from the Plataforma de Integración de
Expedientes (PIE) include personal identifiers, crime classifications, and case
disposition information that are protected under federal and state data
protection regulations. Aggregated results, model specifications, and
additional methodological details are available from the corresponding author
upon reasonable request.

\section*{Declaration of Competing Interests}

The authors declare that they have no known competing financial interests or
personal relationships that could have appeared to influence the work reported
in this paper.

\section*{Funding}

This research did not receive any specific grant from funding agencies in the
public, commercial, or not-for-profit sectors. The collaboration between
Tecnológico de Monterrey and the Fiscalía General de Justicia del Estado de
Zacatecas was conducted as part of ongoing institutional cooperation on public
sector innovation.

\section*{Author Contributions (CRediT)}

\textbf{Fernanda Sobrino}: Methodology, Software, Formal analysis, Writing --
original draft, Visualization. \textbf{Adolfo De Unánue}: Conceptualization,
Methodology, Supervision, Writing -- review \& editing.
\textbf{Edgar Hernández}: Data curation, Software.
\textbf{Patricia Villa}: Investigation.
\textbf{Elena Villalobos}: Visualization, Data curation.
\textbf{David Aké}: Project administration.
\textbf{Stephany Cisneros}: Investigation.
\textbf{Cristian Paul Camacho Osnay}: Conceptualization, Resources, Project
administration. \textbf{Armando García Neri}: Conceptualization, Supervision,
Resources. \textbf{Israel Hernández}: Data curation, Resources.

\bibliographystyle{apalike}
\bibliography{references}  






\appendix

\section{Construction of the Historical Base-Rate Baseline}
\label{app:baseline}
This appendix describes how we construct an empirical baseline to contextualize
the performance of the machine learning models. The baseline uses a
non-machine-learning prioritization rule and relies only on historical case
outcomes. We evaluate it under the same rolling temporal framework and outcome
definitions as the predictive models.

At each evaluation week $t$, we identify all open and active cases in the MAT
unit. To estimate historical closure rates, we use all case outcomes recorded
across the prosecutor’s office prior to week $t$. This broader historical sample
allows us to estimate stable baseline closure patterns while preserving the
temporal separation between training and evaluation.

We group cases by their principal crime category as recorded in the
administrative system. For each crime group $k$, we estimate the probability
that a case in that group will be finalized within 6 months. We compute these
probabilities using all historically observed cases in the group with
observation dates strictly earlier than $t$. Each historical label indicates
whether the case was finalized within six months of its observation date.

When a crime group contains limited historical data, we smooth its estimated
closure probability toward the overall historical closure rate using an additive
shrinkage approach. This procedure balances group-specific information with
aggregate historical trends and reduces variance caused by small sample sizes.

At week $t$, we assign each open MAT case the estimated six-month closure
probability corresponding to its crime group. We then rank cases in descending
order of this score, select the top 300 cases, and compute Precision@300 using
the same evaluation windows and outcome definitions as the machine learning
models.

This baseline uses only information available prior to the prediction date and
incorporates no case-level features beyond crime category. It therefore reflects
historical institutional patterns rather than learned predictive relationships,
providing a transparent and interpretable point of comparison for the machine
learning results.

\section{Sensitivity to Alternative Statutory Prescription Thresholds}
\label{app:prescription_bounds}
This appendix reports results using the minimum and maximum statutory
prescription thresholds defined in Section~\ref{sec:methodology}, in addition to
the mean threshold used in the main text. Here we summarize how the number of
potentially prescribed cases in the bottom 1000 changes under the more
permissive ($T_{\min}$) and more conservative ($T_{\max}$) thresholds.

\subsection{Best Model per Algorithm}

Table~\ref{tab:prescription_all_thresholds} reports the mean number of
potentially prescribed cases among the 1000 lowest-ranked files for the
best-performing model of each algorithm, under each of the three thresholds.

\begin{table}[h]
    \caption{Mean number of potentially prescribed cases in the bottom 1{,}000 under alternative statutory thresholds}
    \centering
    \begin{tabular}{lccc}
        \toprule
        \textbf{Model} & $T_{\max}$ & $T_{\text{mean}}$ & $T_{\min}$ \\
        \midrule
        Random Forest             & 109.52 & 318.63 & 651.78 \\
        Extra Trees               & 118.53 & 212.48 & 643.45 \\
        Scaled Logistic Regression  & 95.80  & 212.70 & 663.85 \\
        Dummy Classifier            & 31.28 & 50.95 & 138.78 \\
        \bottomrule
    \end{tabular}
    \label{tab:prescription_all_thresholds}
\end{table}

As expected, the conservative threshold $T_{\max}$ yields the smallest number of
flagged cases (roughly 9--12\% of the bottom 1{,}000), while the permissive
threshold $T_{\min}$ yields the largest (around 64--75\%). The mean threshold
$T_{\text{mean}}$ lies between these extremes, flagging roughly 21--32\% of the
low-priority tail depending on the model. The Random Forest model remains the
most effective at concentrating potentially prescribed files in the bottom
rankings across all three thresholds.

\subsection{Best Random Forest Model Groups}
Table~\ref{tab:prescription_rf_groups} presents the same sensitivity analysis
for the three best Random Forest model groups (IDs 32, 37, and 38).

\begin{table}[h]
    \caption{Mean number of potentially prescribed cases in the bottom 1{,}000 for the three best Random Forest models}
    \centering
    \begin{tabular}{lccc}
        \toprule
        \textbf{RF Model Group} & $T_{\max}$ & $T_{\text{mean}}$ & $T_{\min}$ \\
        \midrule
        RF 1 & 106.49 & 316.07 & 654.75 \\
        RF 2 & 109.52 & 318.63 & 651.78 \\
        RF 3 & 108.92 & 318.83 & 659.27 \\
        \bottomrule
    \end{tabular}
    \label{tab:prescription_rf_groups}
\end{table}

Across these top Random Forest configurations, the number of potentially
prescribed cases is very similar under each threshold, with variation of only a
few cases around the mean-rule value of approximately 318 files. This suggests
that the prescription screening results are robust to reasonable changes in
Random Forest hyperparameters as well as to the choice of statutory threshold
within the legally admissible range.

\section{Results for the Unidad de Imputado Desconocido (UID)}
\label{app:uid}

In addition to the main analysis for the Módulo de Atención Temprana (MAT), we
conducted experiments applying the same modeling framework to the Unidad de
Imputado Desconocido (UID), which handles investigations without identified
suspects. UID cases differ substantively from MAT cases: they typically enter
the system with less information about potential perpetrators, progress more
slowly, and often remain open for extended periods with limited procedural
activity.

For UID, we replicate the MAT setup as closely as possible. We restrict the
prediction sample to cases that are open in UID at each evaluation week, define
the outcome as whether the case finalizes within six months, and use the same
feature construction, rolling temporal evaluation, and model class (Decision
Tree, Random Forest, Extra Trees, Scaled Logistic Regression), together with the
empirical and random baselines.

\begin{figure}[h]
 \centering
 \includegraphics[width=\linewidth]{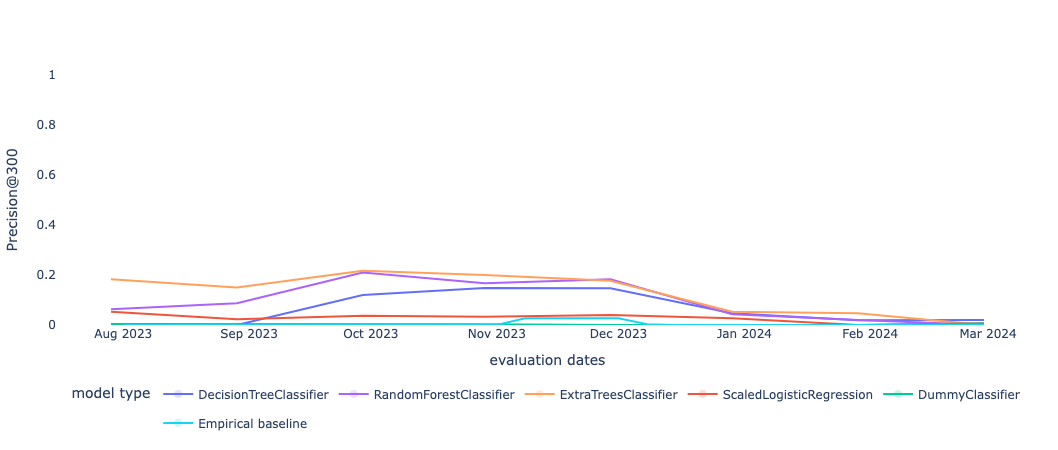}
 \caption{Precision@300 for the best model per algorithm in UID across evaluation periods.}
 \label{fig:uid_precision}
\end{figure}

Figure~\ref{fig:uid_precision} reports Precision@300 over time for the
best-performing model of each algorithm in UID. In contrast to the MAT results,
the models achieve substantially lower precision throughout the evaluation
window, and the gap between the machine learning models and the empirical
baseline is modest. This pattern is consistent with the underlying institutional
dynamics of UID: very few cases without identified suspects resolve within a
short horizon, and the available administrative signals provide limited
information about which of these cases will close next.

Overall, these results suggest that the short-term completion prediction task is
harder in UID than in MAT, and that the current feature set and label definition
provide limited leverage for prioritizing “next-to-finish” cases when suspects
remain unidentified.

We also apply the prescription-screening procedure described in
Section~\ref{sec:methodology} to the UID setting. As in MAT, we examine the 1000
cases with the lowest predicted probability of being concluded within six months
and evaluate them under the mean statutory prescription threshold,
$T_{\text{mean}}$.

\begin{figure}[h]
 \centering
 \includegraphics[width=\linewidth]{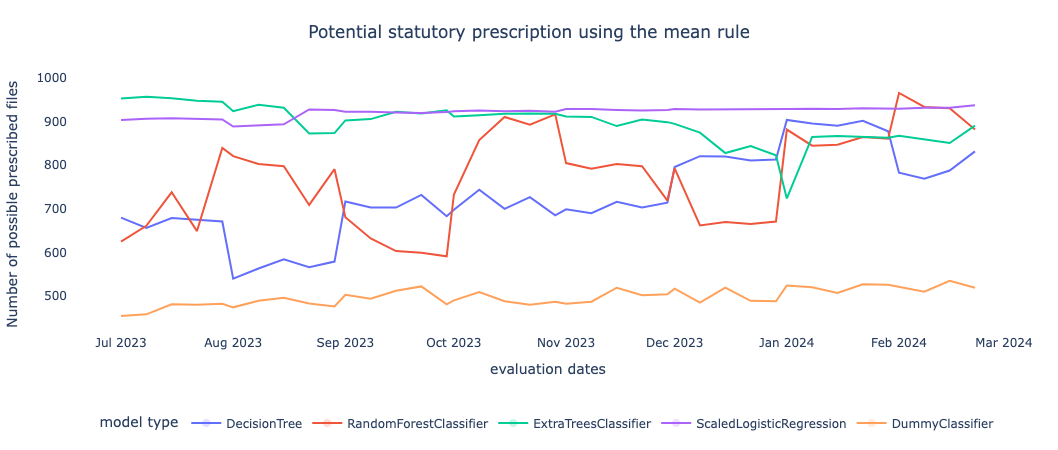}
 \caption{Mean number of potentially prescribed cases among the bottom 1{,}000 UID cases under the mean statutory threshold, by model family.}
 \label{fig:uid_prescription}
\end{figure}

Figure~\ref{fig:uid_prescription} shows the temporal evolution of the mean
number of potentially prescribed cases among these bottom-ranked UID files.
Across all model families, a large share of low-priority UID cases exceeds the
mean statutory threshold, and the number of flagged cases remains high and
relatively stable over time. This pattern reflects the fact that UID
predominantly hosts long-running investigations with limited recent activity,
which are structurally unlikely to progress but may nonetheless remain open in
the administrative system.

To assess the sensitivity of these results to the choice of statutory threshold,
Table~\ref{tab:uid_prescription_thresholds} reports the mean number of
potentially prescribed cases among the bottom 1000 UID files under the mean,
maximum, and minimum prescription rules. For each model, $T_{\text{mean}}$ flags
roughly half to over ninety percent of the lowest-ranked UID cases as
potentially prescribed, while the more conservative $T_{\max}$ flags only around
4--9\% of cases and the more permissive $T_{\min}$ flags roughly 72--97\%.

\begin{table}[h]
    \caption{Mean number of potentially prescribed UID cases in the bottom 1000 under alternative statutory thresholds}
    \centering
    \begin{tabular}{lccc}
        \toprule
        \textbf{Model} & $T_{\text{mean}}$ & $T_{\max}$ & $T_{\min}$ \\
        \midrule
        Decision Tree Classifier      & 734.13 & 77.75  & 863.80 \\
        Random Forest Classifier      & 780.28 & 71.63  & 958.63 \\
        Extra Trees Classifier        & 895.40 & 78.20  & 972.73 \\
        Scaled Logistic Regression    & 922.28 & 92.13  & 954.18 \\
        Dummy Classifier              & 498.53 & 37.30  & 723.40 \\
        \bottomrule
    \end{tabular}
    \label{tab:uid_prescription_thresholds}
\end{table}

Taken together, the UID results indicate that while predictive models add little
value for identifying short-term completions in this unit, the
prescription-screening component remains informative. Even when discriminative
performance on the “next-to-finish” label is weak, the combination of long case
durations and low predicted resolution probabilities yields a concentrated set
of UID files that merit administrative and legal review for potential statutory
prescription. In this sense, the prescription module provides a useful
diagnostic tool for backlog governance in UID, even if prioritization for
near-term resolution is less promising than in MAT.

\end{document}